\documentstyle[prl,aps,floats,epsfig]{revtex}
\newcommand{\be}{\begin{equation}}
\newcommand{\ee}{\end{equation}}
\newcommand{\bees}{\begin{eqnarray}}
\newcommand{\ees}{\end{eqnarray}}
\newcommand{\ra}{\rightarrow}

\newcommand{\bpsi}{{\bar{\psi}}}
\newcommand{\bphi}{{\bar{\phi}}}
\newcommand{\blam}{{\bar{\lambda}}}

\begin{document}

\par
\begingroup

\begin{flushright}
 IFUP-TH 18/98\\
 April 1998\\
\end{flushright}

{\large\bf\centering\ignorespaces
Supersymmetric Vacuum Configurations in String Cosmology
\vskip2.5pt}
{\dimen0=-\prevdepth \advance\dimen0 by23pt
\nointerlineskip \rm\centering
\vrule height\dimen0 width0pt\relax\ignorespaces
Stefano Foffa$^a$, Michele Maggiore$^a$ and Riccardo Sturani$^{b,a}$
\par}
{\small\it\centering\ignorespaces

(a) Dipartimento di Fisica, via Buonarroti, 
I-56100 Pisa, Italy, and  INFN, sezione di Pisa.\\
(b) Scuola Normale Superiore, piazza dei Cavalieri 7,I-56125 Pisa.
\par}

\par
\bgroup
\leftskip=0.10753\textwidth \rightskip\leftskip
\dimen0=-\prevdepth \advance\dimen0 by17.5pt \nointerlineskip
\small\vrule width 0pt height\dimen0 \relax

We examine in a cosmological context the conditions for unbroken
supersymmetry in $N=1$ supergravity in $D=10$ dimensions. We show that the
cosmological solutions of the equations of motion
obtained considering only the bosonic sector correspond to
vacuum states with spontaneous supersymmetry breaking. With a non vanishing
gravitino-dilatino condensate we find a solution of the equations of
motion that satisfies
necessary conditions for unbroken supersymmetry and that smoothly
interpolates between  Minkowski space and DeSitter space with a
linearly growing dilaton, thus  providing a possible example of a
supersymmetric and non-singular  pre-big-bang cosmology. 

\par\egroup
\vskip2pc
\thispagestyle{plain}
\endgroup
Supersymmetric vacuum states in string theory have
 been searched restricting to time-independent
fields in the low-energy
effective action. The result obtained in this case 
is well known~\cite{CHSW}: looking for a
vacuum state of the form $T^4\times K$, where $T^4$ is a maximally
symmetric four dimensional space and $K$ a compact six manifold, one
finds that $T^4$ is necessarily Minkowski space, and requiring $N=1$
supersymmetry in $D=4$ space-time dimensions, 
$K$ is found to be a manifold of $SU(3)$
holonomy. In this Letter we address some aspects of
the problem in a cosmological
context, i.e. including a time dependence in the metric, dilaton
field, etc., and we will compare our results with the 
`pre-big-bang' cosmological model proposed in ref.~\cite{GV},
which is based on the
bosonic part of the string effective action.

We start from the action of $N=1$ supergravity in
$D=10$. We perform some simple field redefinitions on the action
presented in ref.~\cite{BRWN} in order to bring it into the so-called string
frame, where it reads\footnote{Our notations are as follows:
$\phi$ is the dilaton field, $H_{MNP}$ is the field strength of
the two-form field $B_{MN}$,  the graviton
$\psi_M$  is a left-handed Weyl-Majorana spinor, the dilatino
$\lambda$ is a right-handed Weyl-Majorana spinor, and the covariant
derivative $D_M$ is with respect to the spin connection $\omega (e)$,
which is independent of the fermionic fields~\cite{vN,BRWN}.
Indices $A,B,M,N$ take values  $0,\dots ,9$. We use
the signature $\eta_{MN}=(-,+,+,\ldots ,+)$. The conventions for the
Riemann and Ricci tensors are 
$R^{M}_{\hspace*{2mm}NRS}=\partial_R\Gamma^M_{NS}-\ldots$,
$R_{NS}=R^{M}_{\hspace*{2mm}NMS}$; $\Gamma^M_{NS}$ is the Christoffel
symbol while $\gamma^{ABC\ldots}$ denotes the antisymmetrized product
of ten-dimensional gamma matrices, with weight one. }
\bees\label{act}
S&=&\frac{1}{2\kappa^2}\int d^{10}x\sqrt{-g}e^{-\phi}
\left[ R+(\partial\phi )^2-\frac{1}{12}H_{MNP}H^{MNP}
-\bar{\psi}_M\gamma^{MNP}D_N\psi_P -
\bar{\lambda}\gamma^MD_M\lambda
-\frac{1}{2\sqrt{2}}(\partial_N\phi )
\bar{\psi}_M\gamma^N\gamma^M\lambda 
 \right.\nonumber\\
& &\left. +(\partial_N\phi )\bar{\psi}^N\gamma^M\psi_M 
-\frac{1}{32}(\bar{\lambda}\gamma^{ABC}\lambda )
\left(\frac{1}{12}\bpsi^M\gamma_{ABC}\psi_M+
 \frac{1}{2}\bpsi^M\gamma_{MAB}\psi_C -\bpsi_A\gamma_B\psi_C\right)
+\ldots\right]\, .
\ees
The dots in
eq.~(\ref{act}) are terms of the type $(H_{MNP}\times$ fermion
bilinears), terms with three gravitino fields and one dilatino, and
terms with four gravitinos. Their explicit form is not needed below
and can be obtained from ref.~\cite{BRWN}.

If $|\Omega\rangle$ is a vacuum state annihilated by a supersymmetry
generator $Q$, and $\delta\Psi$ is the
 supersymmetry variation of any fermionic field $\Psi$, then
$\langle\delta\Psi \rangle \equiv
\langle\Omega|\delta\Psi |\Omega\rangle =
\langle\Omega|\{Q,\Psi \} |\Omega\rangle =0$.
The variation of the bosonic fields give fermionic fields, and their
expectation values are automatically zero. So, in our case we need to
impose the conditions $\langle\delta\lambda\rangle =
\langle\delta\psi_M\rangle =0$. We consider first the case in which
the expectation values of all bilinears in the Fermi fields are set to
zero. This corresponds to solutions of the equations of motions of the
bosonic part of the action~(\ref{act}), and therefore to the
pre-big-bang cosmology studied in~\cite{GV}. The supersymmetry
variations of the dilatino and gravitino field can be found, e.g., in
ref.~\cite{BRWN}. Writing them in the string frame, and setting
the fermion condensates to zero, the equations 
$\langle\delta\lambda\rangle = \langle\delta\psi_M\rangle =0$
give
\be\label{eq1}
\left(\gamma^M\partial_M\phi -\frac{1}{6}H_{MNP}\gamma^{MNP}\right)\eta =0
\ee
\be\label{eq2}
D_M\eta -\frac{1}{8}H_M\eta =0\, ,
\ee
where $\eta$ is the parameter of the supersymmetry transformation and
$H_M\equiv H_{MNP}\gamma^{NP}$. Note that in the string frame
eq.~(\ref{eq2}) is independent of the dilaton field, contrarily to
what happens in the Einstein frame~\cite{CHSW}. This simplifies
considerably the analysis of the equations.
Writing $\hat{D}_M\equiv D_M-(1/8)H_M$,
eq.~(\ref{eq2}) implies the integrability conditions $[\hat{D}_M,
\hat{D}_N]\eta =0$, which gives
\be\label{int}
\left[
2R_{MNPQ}\gamma^{PQ}+(D_NH_M)-(D_MH_N)-{{H_M}^R}_QH_{NRS}\gamma^{QS}
\right]\eta =0\, ,
\ee
which is therefore a necessary (but not sufficient) condition for
supersymmetry.  One can now see by inspection that the
solutions used in homogeneous 
pre-big-bang cosmology~\cite{GV} do not satisfy 
equations~(\ref{eq1}) and (\ref{int}).
This is obvious for the solutions with vanishing $H_{MNP}$,
since in this case eq.~(\ref{eq1})  requires
a constant dilaton. In fact, we also tried a rather general ansatz
compatible with a maximally symmetric three-dimensional space,
$ds^2=-dt^2+a^2(t)d\vec{x}^2+g_{mn}(t,\vec{y})dy^mdy^n$,
in which the 3-space, with coordinates $\vec{x}$, is isotropic and has 
a scale factor independent of the internal coordinates $\vec{y}$, while
the metric in the six-dimensional internal space is independent of the
$x_i$ but otherwise arbitrary. For $H_{MNP}$ we made the ansatz
$H_{ijk}={\rm const.}\epsilon_{ijk}$ for $i,j,k=1,2,3$,
$H_{0ij}=0$, $H_{MNP}$ vanishes also if indices of the three-space
and indices of the internal space apppear simultaneously,  
and $H_{MNP}$ is arbitrary if all the indices $MNP$ take values
$0,4,\ldots 9$. We also considered  the case of
spatially curved sections of the three-space.
Even with this  ansatz, which is the most general compatible with
maximal symmetry of the three-space when the metric of the three-space
is independent of the internal coordinates, it is straightforward
to show that eqs.~(\ref{eq1},\ref{int}) do not admit  non-trivial
cosmological  solutions.

The super-inflationary pre-big-bang solutions of ref.~\cite{GV} are
therefore rotated by supersymmetry transformations into different
classical solutions of the action~(\ref{act}). 
Since each classical solution of the equations of motion
corresponds to a string vacuum, this means that selecting
such a vacuum  corresponds to a spontaneous  breaking of supersymmetry.
If we want to preserve the advantages of supersymmetry for low-energy
physics, for instance for the hierarchy problem, we do not wish to
break supersymmetry already in the pre-big-bang era. Therefore we now
look  for vacuum states with unbroken supersymmetry. The above results
suggest that, in order to find supersymmetric solutions, the effect
of Fermi fields must be switched on, which means that we must consider
the effect of non-vanishing fermion condensates. We have found a
particularly simple and appealing solution assuming that the only
non-vanishing fermion bilinear is the mixed gravitino-dilatino
condensate, $\langle\blam\psi_M\rangle$. Let us define 
$v_M=-(\sqrt{2}/8)\langle\blam\psi_M\rangle$. It is a composite vector
field and in general in a cosmological setting it depends on 
time (note that while in global supersymmetry the 
fermion condensates are space-time independent~\cite{sQCD}, this is
not the case with local supersymmetry). Furthermore, we look for
solutions with $H_{MNP}=0$.  In this case the equations
$\langle\delta\lambda\rangle =0$, $\langle\delta\psi_M\rangle =0$ give
\be\label{eq3}
\gamma^M(\partial_M\phi -8v_M)\eta =0\, ,
\ee
\be\label{eq4}
D_M\eta -\left(
8v_M+\frac{1}{2}{\gamma_M}^Nv_N\right)\eta =0\, .
\ee
The integrability condition of eq.~(\ref{eq4}) is
\be\label{int2}
\left[ {R_{MN}}^{PQ}\gamma_{PQ} -2v_Av_B 
(g^{AB}\gamma_{MN}+\delta^A_N{\gamma^B}_M- 
\delta^B_M{\gamma^A}_N)-32f_{MN}
+2 ({\gamma_M}^AD_Nv_A-{\gamma_N}^AD_Mv_A)
\right] \eta =0\, ,
\ee
where $f_{MN}=\partial_Mv_N-\partial_Nv_M$ and
$D_Mv_A=\partial_Mv_A-\Gamma_{MA}^Bv_B$.
For the metric we make an isotropic ansatz,
$ds^2=-dt^2+a^2(t)dx_idx^i$, with $i=1,\ldots ,9$, and we define 
as usual the
Hubble parameter $H(t)=\dot{a}/a$. Our strategy is to find a field
configuration $\phi (t),H(t),v_M(t)$ such that 
eqs.~(\ref{eq3}) and (\ref{int2}) are identically satisfied, without
requiring any condition on $\eta$. This is because eq.~(\ref{int2}) is
only the integrability condition for eq.~(\ref{eq4}), and as such 
it is a necessary but not sufficient condition for unbroken
supersymmetry. If it is satisfied for any $\eta$ we still have the
freedom to choose $\eta$ so that also
 eq.~(\ref{eq4}) is satisfied.

Examining eqs.~(\ref{eq3}) and (\ref{int2}) we see that this is possible
only if $v_i(t)=0$, $i=1,..9$. Denoting 
$v_{M=0}(t)\equiv\sigma (t)$,
eq.~(\ref{eq3}) becomes simply $\dot{\phi}=8\sigma$. Eq.~(\ref{int2}),
for $M=0,N=i$, becomes $\dot{H}-\dot{\sigma}+H(H-\sigma)=0$, while
for $M=i,N=j$ we get $(H-\sigma)^2=0$. We see that these equations are
identically satisfied by $H(t)=\sigma (t)$. 
We now ask whether  $H(t)=\sigma (t)$,
$\dot{\phi}(t)=8\sigma (t)$ is a solution of the equations of motions,
as we expect for a supersymmetric configuration.
The equations of motion obtained with a variation with respect to
fermionic fields are automatically satisfied when we take the
expectation value over the vacuum, and we only need to
check the variation with respect to bosonic fields. We
introduce the
shifted dilaton $\bar{\phi}=\phi -d\beta$, where $\beta =\log a$
and $d=9$ is the number of spatial dimensions, and we also retain the
lapse function $N$ in the metric, so that
$ds^2=-N^2dt^2+e^{2\beta}dx_idx^i$.
Restricting to homogeneous fields, the
relevant part of the action can be written as
\be\label{act2}
S=-\frac{1}{2\kappa^2}\int dt e^{-\bar{\phi}}\frac{1}{N}
\left[ -d\dot{\beta}^2+\dot{\bphi}^2+
2\left( -\frac{\sqrt{2}}{8}\blam\psi_0\right) \dot{\bphi}
+2d\left( -\frac{\sqrt{2}}{8}\blam\psi_0\right) \dot{\beta}
-8\left( \frac{\sqrt{2}}{8}\blam\psi_0\right)^2\right]\, .
\ee
The last term in the action~(\ref{act2}) comes from the
term $(\bar{\lambda}\gamma^{ABC}\lambda )(\bpsi^M\gamma_{ABC}\psi_M)$
in eq.~(\ref{act}), making use of the Fierz identity 
$(\bar{\lambda}\gamma^{ABC}\lambda )(\bpsi^M\gamma_{ABC}\psi_M)
=96(\blam\psi^M)(\blam\psi_M)$ (see e.g. the appendix of
ref.~\cite{BRWN}). Instead, the terms 
$(\bar{\lambda}\gamma^{ABC}\lambda )
(\bpsi^M\gamma_{MAB}\psi_C)$ and 
$(\bar{\lambda}\gamma^{ABC}\lambda )
(\bpsi_A\gamma_B\psi_C)$
in the action~(\ref{act}) are independent from 
$(\blam\psi^M)(\blam\psi_M)$ and their condensates can be
consistently set to zero.
We now variate the action with respect to
$N,\bphi ,\beta$ and then take the expectation value of the 
terms $\blam\psi_0$, $(\blam\psi_0)^2$
over the vacuum. We get the equations
\be
\frac{d}{dt}\left( e^{-\bphi}(H-\sigma )\right)=0
\, ,\hspace{10mm}
\dot{\bphi}^2 -9H^2+
2\sigma\dot{\bphi}+18\sigma H
-8\langle (\frac{\sqrt{2}}{8}\blam\psi_0)^2\rangle=0
\ee
\be
2(\ddot{\bphi}+\dot{\sigma})-2\dot{\bphi}(\dot{\bphi}+\sigma )
-9H^2+\dot{\bphi}^2+2\sigma\dot{\bphi}+18\sigma H
-8 \langle (\frac{\sqrt{2}}{8}\blam\psi_0)^2\rangle=0\, .
\ee
For  the configuration $H=\sigma$, $\dot{\phi}=8\sigma$ (and therefore
$\dot{\bphi}=\dot{\phi}-9H=-\sigma$) we see that the equations of
motion are identically satisfied if
$\langle (\blam\psi_0)^2\rangle=
\langle \blam\psi_0\rangle^2$. 
Consistency therefore requires that supersymmetry enforces this
relation between the condensates. In general, it is well known that
relations of this kind are indeed enforced by supersymmetry;
for instance, the  relation $|\langle \bar{\chi}\chi\rangle|^2=
\langle |\bar{\chi}\chi |^2\rangle$ holds
for the gaugino condensate
in the case of super-Yang-Mills theories~\cite{sQCD} and 
in  supergravity coupled to super-Yang-Mills~\cite{DRSW}.

\begin{figure}[t]
\centering
\epsfig{file=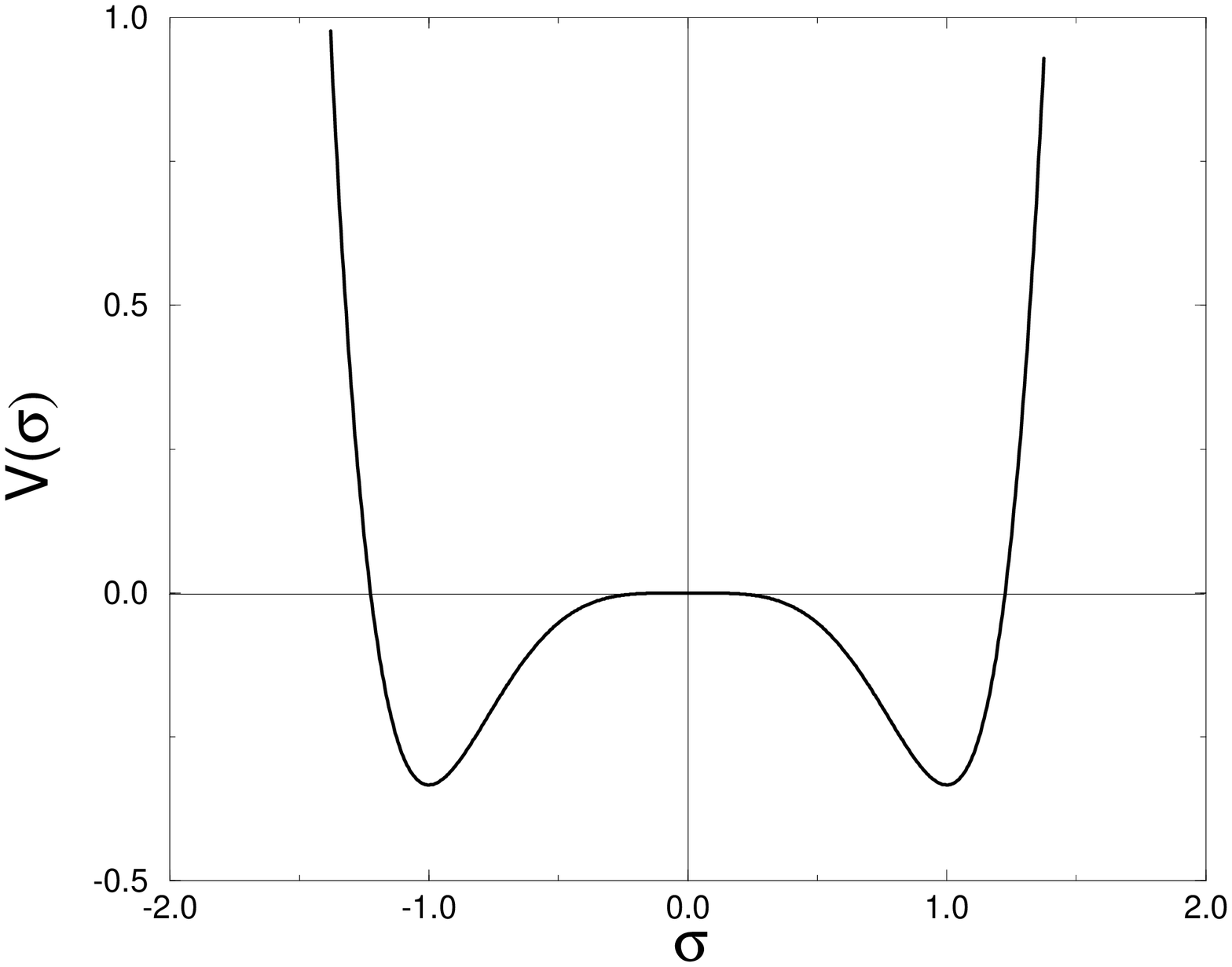,width=0.4\linewidth,angle=270}
\epsfig{file=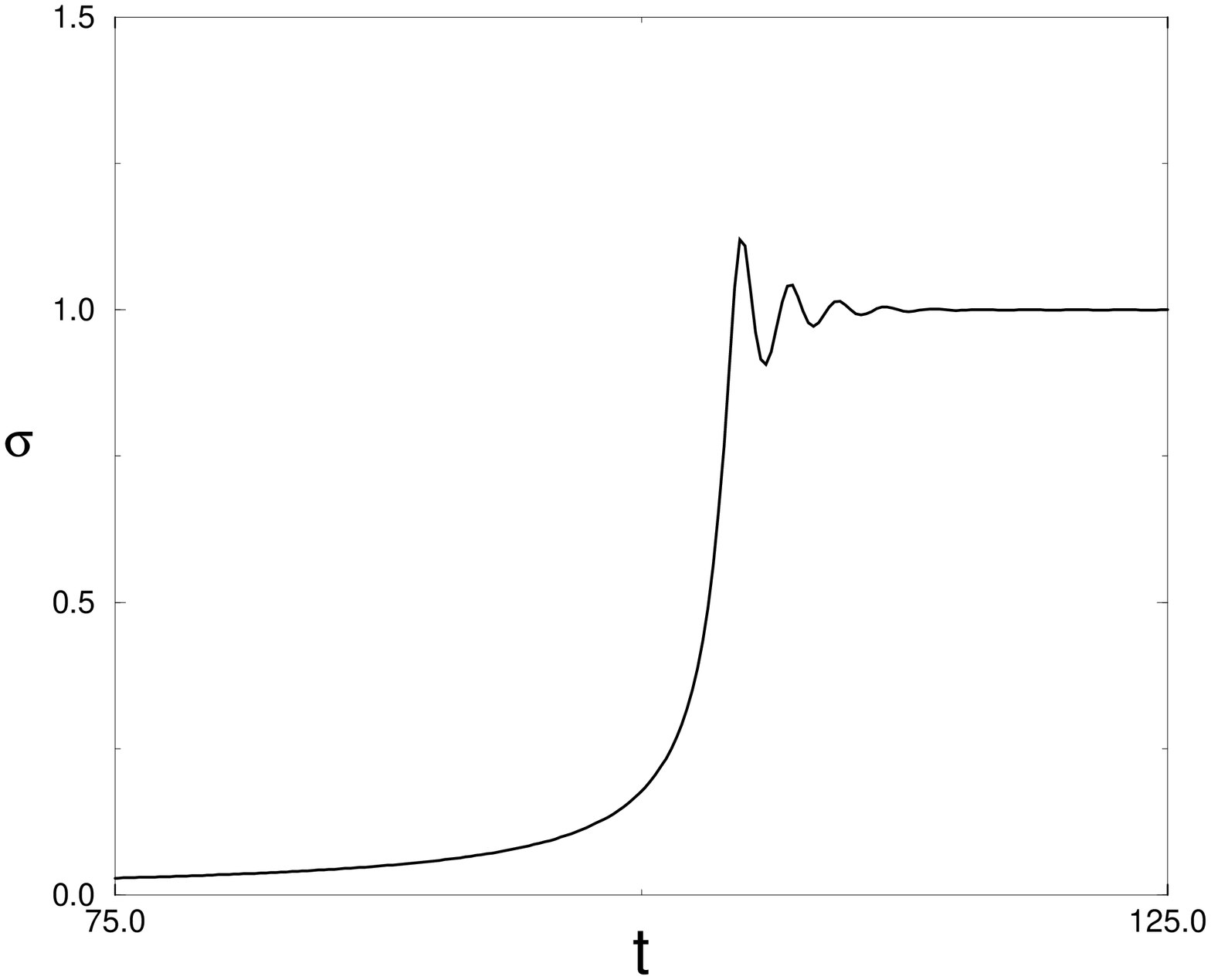,width=0.4\linewidth,angle=270}
\caption{
A symmetry breaking potential for the field $\sigma (t)$.
\hspace*{16mm}FIG.~2. The evolution of the field $\sigma (t)$.}
\end{figure}

It remains to discuss the dynamics of the condensate $\sigma
(t)$. This is a composite field whose dynamics will be governed by an
effective action which in principle follows from the
fundamental action~(\ref{act}). To assume that a condensate forms is the
same as assuming that the field $\sigma (t)$ has an effective action
with a potential $V(\sigma )$ with the absolute minimum 
at $\sigma =\bar{\sigma}\neq 0$, see fig.~1. If we choose as initial
condition $\sigma\ra 0^+$ as $t\ra -\infty$, $\sigma (t)$ will evolve
from $\sigma =0$ toward the positive minimum of the potential, and it will
make damped oscillations around $\sigma =\bar{\sigma}$, the damping
mechanism being provided by the expansion of the Universe and possibly
by the creation of particles coupled to the $\sigma$ field. The
qualitative behaviour of $\sigma (t)$ will be therefore of the form
plotted in fig.~2. For illustrative purposes, we have shown in fig.~2
the evolution of $\sigma$  obtained assuming an effective action, in
the string frame, of the form
\be
S=\int dt\, e^{ -\bphi } \left(\frac{1}{2}\dot{\sigma}^2-V(\sigma )\right)
\ee 
where $V(\sigma )=-\sigma^4+(2/3)\sigma^6$ is the potential shown in
fig.~1.  (We use units such that the minimum is at $\bar{\sigma}=1$.)
This gives the equation of
motion $\ddot{\sigma}-\dot{\bphi}\dot{\sigma}+V'=0$, with 
$-\dot{\bphi}=\sigma$ providing the friction term. However, the
qualitative behaviour is independent from these specific choices.

Since $H(t)=\sigma (t)$ and  
$\dot{\phi}=8\sigma (t)$, this solution 
corresponds to a cosmological model that starts at $t\ra -\infty$
from Minkowski
space with constant dilaton and vanishing fermion condensates,
i.e. from the string perturbative vacuum, and evolves toward a 
De~Sitter metric $H=$ const., with linearly growing dilaton. This is
similar to the scenario found in ref.~\cite{GMV} in the bosonic sector with
$\alpha '$ corrections. In the present case the scale at which the
curvature is regularized is given by the fermion condensate 
$\bar{\sigma}$ while in \cite{GMV} it was given by the $\alpha '$
corrections. However, in the case studied in ref.~\cite{GMV}, 
higher order  $\alpha '$ corrections were not  under control, so that a
definite statement about the effectiveness of the regularization
mechanism could not be made (see also the discussion in
ref.~\cite{MM}). In the present case, instead, the fact that $\sigma$,
and therefore $H$ and $\dot{\phi}$, stops growing follows from the general
requirement that the potential $V(\sigma )$ be bounded from below 
and has a minimum, as
we  expect for the effective potential derived from any well-defined
fundamental action as the action~(\ref{act}). As in the case studied
in~\cite{GMV}, the DeSitter solution should finally be matched to a
standard radiation-dominated era. For the matching, $O(e^{\phi})$
corrections to the string effective action
are probably important~\cite{BM}, 
since $\dot{\phi}$ is positive and therefore
at some stage  $e^{\phi}$ becomes large. 
 The so-called graceful exit
problem however takes a different form in our scenario, since 
$\dot{\bphi}=-\sigma$ is always negative in our model and no `branch
change'~\cite{BV} needs to occur. Instead,
when the gauge coupling
$\sim e^{\phi}$ becomes strong, gaugino condensation is also expected
to occur, suggesting that the gaugino condensate might play  a role in 
matching the De~Sitter phase to a radiation dominated era.

It is also interesting to observe that, if at small $\sigma$ the potential 
$V(\sigma )$ behaves as $V(\sigma )\simeq -\sigma^4/(2c^2)$, with 
$c$ a positive constant, then at large negative values of time the
solution of the equation of motion for $\sigma$ is $\sigma
(t)\simeq c/(-t)$ and therefore $H(t)\simeq c/(-t)$, corresponding to a
super-inflationary  stage of expansion, as it also happens for the
solutions found in~\cite{GV} in the bosonic sector of the model.

We conclude stressing that the solution that we have presented has more an
illustrative rather than a realistic value. For one thing, we have
discussed an isotropic ten-dimensional solution and we have not
touched upon the issue of compactification of the extra dimensions. To
obtain an anisotropic cosmological model, in which three spatial
dimensions expand and six get compactified, most probably one must
switch on the effect of the three-form $H_{ABC}$~\cite{anis} 
or include the effect
of a  dilatino condensate $\blam\gamma_{ABC}\lambda$, or a gravitino
condensate such as $\bpsi_A\gamma_B\psi_C$, or a gaugino condensate
$\bar{\chi}\gamma_{ABC}\chi$. They can separate 3 spatial dimensions
from the remaining six if either all the three indices $A,B,C$ belong
to the three dimensional space, or if they are the holomorfic indices
of a six-dimensional complex manifold.
Still, the toy solution that we have discussed illustrates the
 role of fermion condensates in a supersymmetric cosmology and
provides a novel mechanism for the regularization of the singularity
of the pre-big-bang cosmological solutions.

\end{document}